# Measurement of ultrashort vector pulses from polarization gates by in-line, single-channel spectral interferometry

Benjamín Alonso[1,2,†] and Íñigo J. Sola[2,‡]

[1]Sphere Ultrafast Photonics, S.A., Parque de Ciência e Tecnologia da Universidade do Porto, R. do Campo Alegre 1021, Edifício FC6, 4169-007 Porto, Portugal

[2]Grupo de Investigación en Aplicaciones del Láser y Fotónica, Departamento de Física Aplicada, Universidad de Salamanca, Salamanca, E-37008, Spain

[†] E-mail: b.alonso@usal.es
[‡] E-mail: ijsola@usal.es

*Abstract*. The growing use of ultrashort laser pulses exhibiting time-varying polarization (vector pulses) demands simple and robust characterization techniques capable to perform measurements in a broad range of experimental and environmental conditions. Here we present in-line, single-channel setup based on spectral interferometry to characterize ultrashort vector pulses. The use of a *bulk* interferometer based on birefringence is key for the stability and sensitivity of the technique, thus being simple and highly robust. The technique is used to measure vector pulses corresponding to polarization gates, which are used in many applications. Those results are validated by simulations. The technique here presented has a number of potential applications in nonlinear effects (e.g. transient birefringence and nonlinear phenomena with vector pulses).

## 1. INTRODUCTION

Advances on laser technology are enabling the generation and use of pulses of light more and more complex, both on time (e.g., single-cycle pulses) and singular spatial distributions (e.g., ultrashort pulsed vortex, Bessel beams, etc.). In addition, there is a raising interest on ultrashort pulses exhibiting time-varying polarization (vector pulses) [1-4]. The use and analysis of vector pulses is a useful tool in the study of quantum wells properties [5], vector coherent control for selective isomerization of enantiomers [6-9] and the generation of THz pulses with time evolving polarization from IR vector pulses [7], to mention some examples.

In the *attoscience* field, vector pulses are used to alter the high-order harmonic generation (HHG) process in gas. Because of the high sensitivity of the HHG on atoms to light polarization (when light polarization is not linear, HHG efficiency vanishes), ultrashort pulses with particular temporal evolving polarization, known as polarization gates (PG), are used to control the number of attosecond light bursts created in the process [10-12]. When applied to few-cycle pulses, the resulting vector pulses from the PG are able to generate isolated attosecond pulses through HHG [13]. In a different configuration, two noncollinear counter-rotating circularly polarized beams create a vector pulse that generates isolated circularly polarized attosecond pulses by HHG [14].

In this context, the characterization of the time-dependent polarization of the pulse is a key point. The reconstruction of ultrashort pulses has been developed since decades. Nowadays, techniques such as FROG [15], SPIDER [16] or d-scan [17], are well established and allow to measure pulses down to few optical cycles and single-cycle regime [18–22]. However, the usual techniques are designed for scalar pulses where pulse polarization is linear and constant on time.

During the last years, different approaches have been proposed to accomplish vector pulse reconstruction in the femtosecond range, e.g. through spectral interferometry in different configurations [23,24], tomographic reconstruction from measurement of different projections [25], time-resolved ellipsometry [26], some variants [27,28] based on X-FROG technique [29], nanointerferometric measurement of vector beams [30], or spatial-spectral interferometry [31].In particular, in dual-channel spectral interferometry method, known as POLLIWOG [21], a linearly polarized known reference pulse interferes with a delayed test (unknown) pulse.

By using spectral interferometry analysis [32], the relative phase between two polarization components of the test pulse is extracted. Together with the amplitudes for each polarization projection and the spectral phase of the reference pulse (measured with any scalar temporal pulse reconstruction technique), the time-dependent polarization pulses can be reconstructed. The use of a linear process as interferometry makes it a sensitive and powerful technique, with a non-iterative retrieval algorithm. However, precisely because of this interferometric nature, typical schemes commonly based on standard interferometers such as Mach-Zehnder or Michelson will be very demanding on stability, in order to preserve the spectral fringes and, more important, to accurately measure the phase difference between the orthogonal components. In fact, the POLLIWOG uses a dual-channel spectrometer to record separately and simultaneously the interferences of the x and y-components, thus preventing shot-to-shot phase fluctuations to affect the polarization component relative phase measurements. Therefore, it results that, to determine the time-varying polarization state, it is a key point the accurate measurement of the relative amplitude and, primarily, the phase of the polarization component. In this context, to fulfill the stability requirements, our proposal is based on an in-line single-channel spectral interferometry setup, in which the relative phase is directly measured in a crossed intermediate angle projection. The objective of the present work is to implement a robust and reliable technique, not affected by instabilities and noise, and simple in order to be applied to different experimental cases, e.g. PGs that are used for different applications.

Firstly, we present the description of the method and the set-up used for the experimental implementation, accompanied with a first experimental validation of a vector pulse and a study of stability. Secondly, we apply it to the reconstruction of vector pulses obtained from PG technique. All measurements are corroborated by simulations. Finally, we summarize the conclusions.

## 2. THE METHOD: IN-LINE SPECTRAL INTERFEROMETRY

Our method takes the advantages of using interferometry, a process characterized by being linear, high sensitivity or a direct analysis, while avoiding the main drawback, i.e., the instabilities and alignment requirements. Mostly, spectral interferometry set-ups are based on interferometers, where the unknown pulse is combined with the one serving as a reference. These setups are very sensitive to vibrations and instabilities from the environment. To prevent this problem, we use a birefringent element as a monolithic in-line interferometer, a calcite plate (3 mm of thickness), with their optical axes parallel to the input and output faces. When an incoming vector pulse passes through the plate, it is split into the ordinary and extraordinary components, presenting a delay due to the different refractive indices of the birefringent material (in our case, 1.8 ps).

In previous works, birefringent crystals have already been used to create stable delayed pulse replicas [33-35], which has been applied to measure the spectral phase of scalar pulses [34-37].

After the calcite plate, we use a linear polarizer (LP) to select different polarization projections of the delayed components and the corresponding spectra are subsequently acquired by a spectrometer (AvaSpec 2048-USB1, from Avantes Inc.). Just by varying the LP orientation, it is possible to measure the spectra from the ordinary and extraordinary components respectively and the interference between both at an intermediate angle. This latter spectrum encodes the spectral phase difference among the ordinary and extraordinary components (once the phase added by the birefringent plate is subtracted). Therefore, a vector component of the pulse is actually acting as a reference for the orthogonal one. The use of a *bulk* interferometer means that both components travel through the same physical path and there is not any other pulse splitting or recombination, thus awarding the pursued stability to the detection, as well as getting rid of a precise alignment. Finally, if one of the components is characterized and its spectral phase is known (e.g., by means of a standard scalar temporal pulse reconstruction technique), it is possible to reconstruct the vector pulse by applying Fourier-transform spectral interferometry analysis (FTSI) [32] to the intermediate projection.

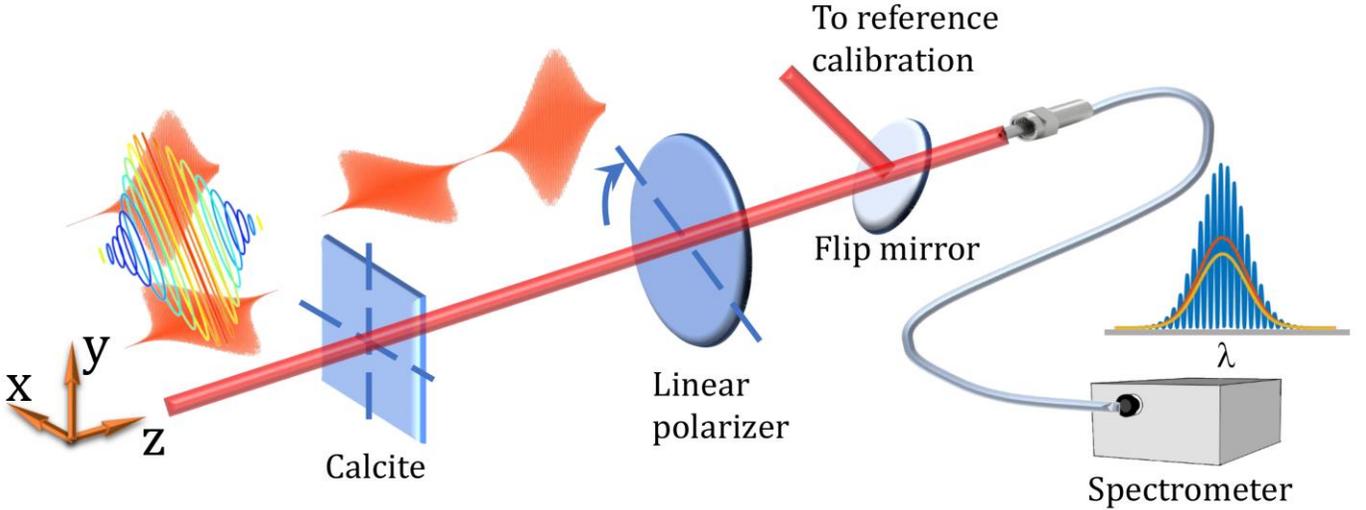

Fig. 1. Schematic of the detection setup for the polarization measurements. The vector pulse passes through a birefringent material (calcite) where the x- and y-components are delayed because of the different dispersion. A linear polarizer selects different projections to be measured in the spectrometer (horizontal and vertical projections for the amplitude; intermediate projection to obtain the relative phase of the pulse components). The spectral phase of the horizontal component is measured with an auxiliary device. The system is calibrated using a pulse linearly polarized at 45º.

The experimental setup and the axes convention are represented in Fig. 1. Rotation angles for the different elements will be given with respect to the x-axis and being the observer looking to the source (positive rotation angle means counter-clockwise). Firstly, the unknown vector pulse impinges the birefringent plate (calcite 3 mm thickness, 0.5"-side square) that is located with the extraordinary (fast) axis horizontal (x-axis), being the y-component of the pulse delayed with respect to the x-component. The LP can be situated at different angles $\alpha$, selecting the corresponding pulse projections, $S_\alpha$, that are detected in the fiber coupled spectrometer.

To calibrate in amplitude and phase the system, we create a calibrating pulse linearly polarized at 45º. This pulse has the same spectral amplitudes $A_y(\omega) = A_x(\omega)$ and phases $\phi_y(\omega) = \phi_x(\omega)$, so we can obtain the amplitude response of the system and the relative dispersion of the calcite axes.

The horizontal and vertical projections of the LP provide the information of the amplitude of the x- and y-components, with $S_{0°} = S_x$ and $S_{90°} = S_y$ respectively. The intermediate projection at $\alpha = 45°$ encodes the relative phase between the orthogonal components (x and y):

$$S_{45°} = \tfrac{1}{2} S_x + \tfrac{1}{2} S_y + \sqrt{S_x S_y} \cos\left[\left(\phi_y - \phi_x\right)_{meas}\right]. \quad (1)$$

This relative phase gathers the pulse phase and the calcite phase, $\left(\phi_y - \phi_x\right)_{meas} = \left(\phi_y - \phi_x\right)_{pulse} + \left(\phi_y - \phi_x\right)_{calc}$, being the calcite contribution the one that introduces the delay required to use the FTSI algorithm. In order to remove the calcite relative dispersion, we used the projection $S_{45°}$ of the calibrating pulse, for which $\left(\phi_y - \phi_x\right)_{pulse}^{[Calibr]} = 0$ and $\left(\phi_y - \phi_x\right)_{meas}^{[Calibr]} = \left(\phi_y - \phi_x\right)_{calc}$.

Regarding the amplitude sensitivity of the fiber coupled spectrometer to the x and y components, it is corrected with the 0º and 90º projections of the same calibrating pulse ($S_y = S_x$). The y-component of the pulse to be measured is corrected as $S_y^{[Pulse]} = S_y^{[Pulse]} R_{calibr}$, where $R_{calibr}(\omega) = S_y^{[Calibr]} / S_x^{[Calibr]}$ is smoothed and interpolated within the spectral range of the calibrating pulse.

The spectral phase of the component acting as a reference, in our case the x-component, is measured with a SPIDER device (any temporal characterization technique is valid). With the LP set horizontal and a flip mirror, we extract this reference pulse. Under this configuration, the x-axis dispersion of the calcite [38] needs to be subtracted (alternatively, the projection of this reference pulse can be extracted before the calcite).

The proposed scheme requires that both the x- and y-components of the pulse have similar spectral content and comparable amplitude in order to obtain suitable spectral interferences. In the case of having LP polarized at 0º or 90º (trivial cases), it is possible to rotate the characterization system or the pulse e.g. 45º to optimize the contrast in the detected fringes (as we do for the large PG case in the next section presenting results). On the other hand, if having different spectra at 0º and 90º, the same rotation would merge the spectra and lead to fringes in the whole bandwidth.

The technique proposed is multi-shot as it uses 3 individual projections of the calibrating and unknown pulses, plus the reference pulse characterization. The common-path configuration prevents that shot-to-shot fluctuations of the interferometer (due to thermal variations, mechanical vibrations or air fluctuations) affect the measurement. Laser pulse energy fluctuations may affect the retrieval of the relative amplitude of the x- and y-components. In the case of very unstable laser systems, this could be circumvented by using dual-channel spectrometers to simultaneously measure two projections, together with the retrieval of the relative amplitude between $S_x$ and $S_y$ directly from the intermediate projection $S_{45°}$, in a similar way than in [32].

## 3. RESULTS: EXPERIMENTS AND SIMULATIONS

### A. Description of the experiments and parameters

For the experiments, we have used a CPA Ti:sapphire laser (Spitfire, from Spectra Physics) at a repetition rate of 1 kHz, emitting 100 fs (FWHM) pulses. To select the different spectral projections, we used an achromatic LP (LPVIS050, from Thorlabs). For the simulations, we have used the actual experimental spectrum of the pulse and then calculated the corresponding polarization shaping for the different cases. Notice that our pulse is centered at 797 nm, while the wave plates employed here are designed for 800 nm.

From the experimental characterization, we obtain the temporal amplitude and phase of the vector pulse, from which we calculate the polarization state as a function of frequency and time, presenting the following magnitudes in the results of Fig. 2,4,5: spectrum (S) and intensity (I) for x and y components, ellipticity $\varepsilon = b/a$ calculated as the minor to major axes ratio, phase difference $\delta$ between the x and y components , and orientation of the polarization ellipse (azimuthal angle $\chi$) referred to the x-axis. All vector pulse measurements have been done 10 times, then calculating the statistical error as the standard deviation and representing it in all the retrieved magnitudes as a gray shaded area (when not seen it means that it is so small that it is not distinguished from the mean value).

### B. Validation and stability study

The performance of the technique has been tested at known cases, trivial and non-trivial. For example, the laser pulses have passed through an achromatic half-wave plate, in order to rotate their constant polarization direction, which has been verified in the measurements. Also, we have characterized the input LP pulse passing through a zero-order quarter-wave plate at 45º creating constant circular polarization.

Then, a first non-trivial validation consisted in the measurement of a pulse linearly polarized at 0º, after passing through a multiple-order quarter-wave plate (QWPM-800-10-4 from CVI, which we label QWPM in the whole manuscript) with the slow axis at 45º. This element consisted of a 652-$\mu m$ thick quartz plate, which introduces a delay of around 20 fs between the fast and slow axes for 800 nm.

We have calibrated it by studying the spectrum transmitted when the plate is placed at 45º between two crossed linear polarizers, using as light source both spectrally broadened pulses and incoherent white light. By rotating the last polarizer, clearly the plate behaves as a quarter-wave plate around 800 nm, half-wave plate at 775 nm and full-wave plate at 825 nm, all at multiple-order regime. Thus, when the light passes through the plate, oriented at 45º, the state of polarization becomes time-varying due to the delay between ordinary and extraordinary components, being not negligible compared to the pulse duration.

To simulate the vector pulse, we have calculated the dispersion introduced by the QWPM thickness imposing the phase difference $\pi/2$ at 800 nm. The QWPM angle used in the simulations is 44º (notice that this is compatible with the experimental error in our rotation mount), being the value for which the simulations match the experimental results ($\varepsilon$, $\delta$, and $\chi$, simultaneously). Fig. 2 shows the vector pulse reconstruction and simulation results. The spectral amplitude x,y components are laterally shifted (due to the effect of dispersion in a multi-order plate), the spectral ellipticity reaches a maximum ($\varepsilon \lesssim 1$) with circular polarization (CP) at 800 nm, while the peak at the ellipticity depending on time (close to circular polarization) is shifted to positive times with respect to the peak intensity (we have corroborated that this is because the angle of 44º instead of 45º).

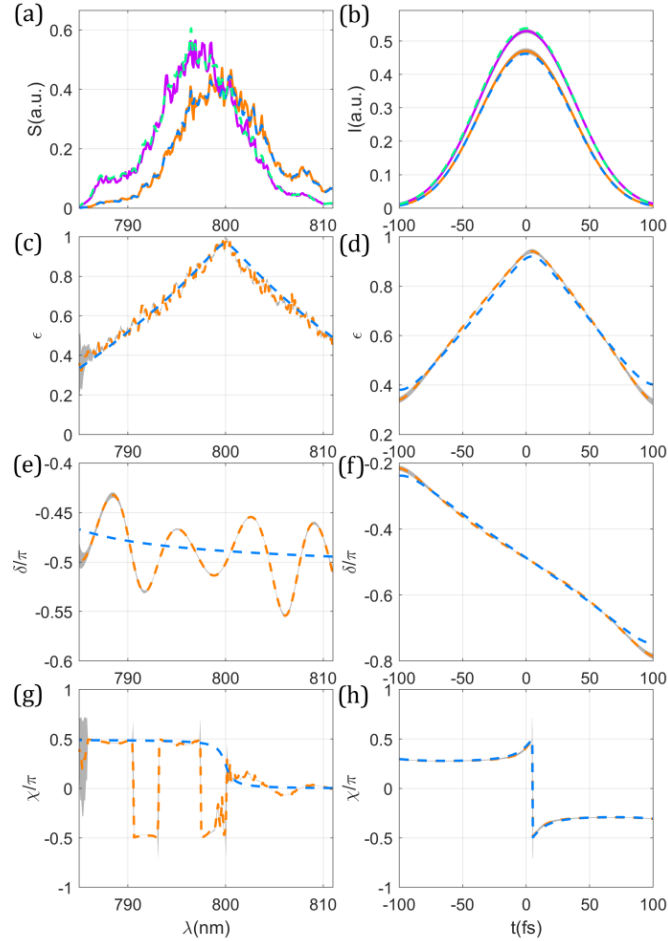

Fig. 2. Polarization results for the case input LP 0º, followed by QWPM slow axis at 45º. Spectral/temporal (a/b) intensity, (c/d) ellipticity $\varepsilon$, (e/f) phase difference $\delta$, and (g/h) azimuth $\chi$, respectively. Plots (a,b), experimental *x* (solid red) and *y* (solid magenta), simulated *x* (dashed blue) and *y* (dashed green) components. Plots (c-h), experimental (dashed red), simulated (dashed blue). In all plots the gray shaded area stands for the experimental error.

In order to show the stability of the technique, we used the interference fringes for a pulse with linear polarization at 45º. We did two stability studies, a short-term one with 300 measurements, 1 per second during 5 minutes, and a long-term one with 1500 measurements, 1 each 5 seconds during $\gtrsim 2$ hours. The fringes where stable and we did not observe the typical mechanical and thermal drift of standard interferometers, as seen in Fig. 3 for the long-term study. Using a fine analysis with the FTSI algorithm, we found the standard deviation of the phase fluctuations to be $9 \cdot 10^{-4} \cdot 2\pi$ rad and $4 \cdot 10^{-3} \cdot 2\pi$ rad for the short and long-term studies, respectively. Those values are well below the typical values in standard and non-standard interferometers [39-42]. The key point for this stability is the in-line geometry that also confers robustness, accuracy and simplicity to the setup. As said before, this is of paramount importance in the case of the determination of the relative phase between the two polarization components, since it is critical for an accurate calculation of the pulse polarization state. Therefore, the setup can be integrated easily and work at environmental conditions less controlled than the present at a laboratory.

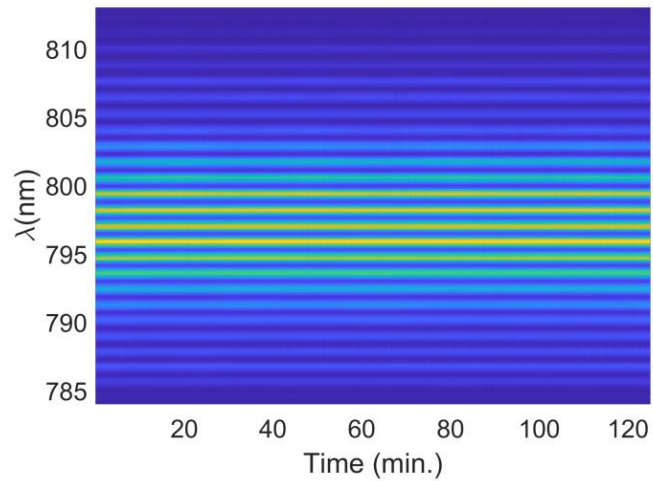

Fig. 3. Spectral interferences recorded with the in-line interferometry setup during 125 minutes used for the long-term stability analysis.

### C. Study of polarization gates

Once the performance of the method was tested, we studied the vector pulse coming from PG pulses. As commented previously, that kind of vector pulses are used as one of the main techniques to obtain isolated attosecond pulses [12,13]. First attempts were performed by using an interferometer in order to delay two orthogonal polarization components of an incoming linearly polarized pulse, in order to obtain circular polarization at the center of the outcoming pulse, and linear polarization at the edges [10]. A zero-order quarter-wave plate (QWP0) aligned at 45º converted the linear polarization into circular and the circular polarization into linear, obtaining a vector pulse exhibiting linear polarization only in its center, while ellipticity rises at the edges. A simpler and more compact version consists in using a combination of a multiple-order (QWPM) and a zero-order (QWP0) quarter-wave plate [10,11]. The element QWPM is the same than before, whereas the element QWP0 is a zero-order quartz plate designed for operation at 800 nm (QWP0-800-08-4-R10 from CVI). The input pulse has LP at 0º, then the QWPM is orientated with the slow axis at 45º. After that, the QWP0 is situated with the fast axis at 90º to create the so-called narrow PG (results in Fig. 4), characterized by having linear polarization at the center of the pulse and circular polarization at the pulse edges.

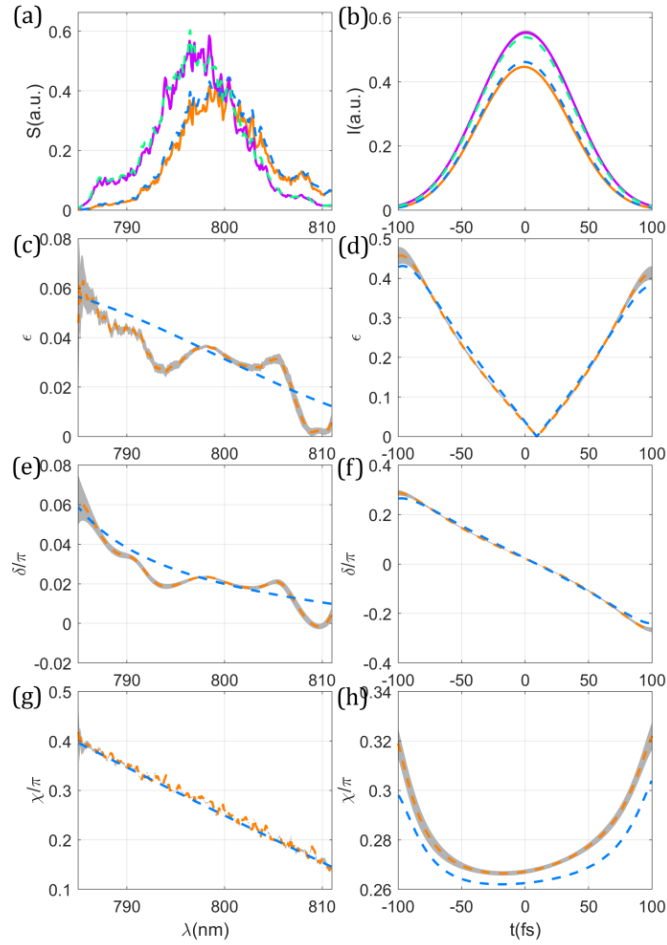

Fig. 4. Polarization results for the narrow PG (input LP 0º, followed by QWPM slow axis at 45º, and QWP0 fast axis at 90º). Spectral/temporal (a/b) intensity, (c/d) ellipticity $\varepsilon$, (e/f) phase difference $\delta$, and (g/h) azimuth $\chi$, respectively. Plots (a,b), experimental *x* (solid red) and *y* (solid magenta), simulated *x* (dashed blue) and *y* (dashed green) components. Plots (c-h), experimental (dashed red), simulated (dashed blue). In all plots the gray shaded area stands for the experimental error.

After the QWPM, the pulse is that presented in Fig. 2, with CP at the maximum intensity, with ellipticity decreasing at the edges of the pulse (ideally reaching LP if the delay introduced between the QWPM components is high enough compared to the pulse duration, which is not our case). If then QWP0 is at 90º (with respect to the x-axis) the CP of the maximum is converted to LP and the ellipticity at the edges now increases (in the limit ideally being CP), thus creating a narrow PG with an effective LP region shorter than the initial pulse duration. The azimuth is almost constant at 45º in the temporal domain, and the direction of rotation is seen to change at the maximum intensity (with the sign of $\delta$).

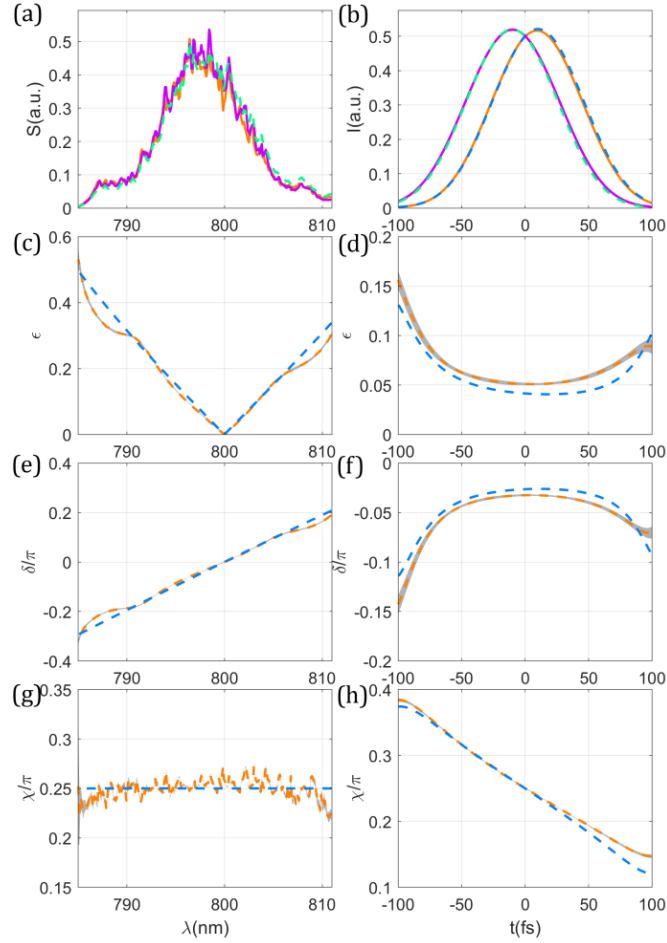

Fig. 5. Polarization results for the large PG (input LP 45º, followed by QWPM fast axis at 0º, and QWP0 fast axis at 90º). Spectral/temporal (a/b) intensity, (c/d) ellipticity $\varepsilon$, (e/f) phase difference $\delta$, and (g/h) azimuth $\chi$, respectively. Plots (a,b), experimental *x* (solid red) and *y* (solid magenta), simulated *x* (dashed blue) and *y* (dashed green) components. Plots (c-h), experimental (dashed red), simulated (dashed blue). In all plots the gray shaded area stands for the experimental error.

In the so-called large PG, the pulse is characterized because the temporal ellipticity is zero (LP) while the azimuth of the LP is rotating along the pulse duration. The fast axes of the QWP0 and QWPM wave plates are aligned at 45º with respect to the input LP. Under this configuration, the output pulse would be almost LP at 90º. To avoid this, we rotate the system 45º, where the initial pulse passes through a LP at 45º (polarizing cube), then the QWPM and the QWP0 wave plates are orientated with their fast axes at 90º (results in Fig. 5). The temporal ellipticity is kept almost constant below 0.1, LP as expected, while the polarization azimuth evolves on time (from 90º to 0º, being 45º at the center of the pulse).

A full experimental characterization of the PGs [43] is important to predict, optimize and understand the applications of those pulses [44], considering that the real pulse may differ from the expected one e.g. due to the use of non-ideal retarders or non-Gaussian pulse spectrum.

## 4. Conclusions

In sum, we have presented a simple and compact in-line, single-channel device to reconstruct vector pulses. The capabilities of the technique have been demonstrated with different cases of known pulses, being the results corroborated by the simulations. The small statistical error in the retrievals shows a good performance in the measurements, which is related to the excellent stability of the compact *bulk* interferometer. We have used it to fully characterize PGs of interest for their associated applications. The robustness of the setup makes it suitable in more demanding future experimental situations, e.g. after nonlinear propagation, out laboratory conditions or to characterize vector beams that are nowadays used in many applications.

## 5. FUNDING

This work was supported in part by Junta de Castilla y León under Grant SA287P18, Spanish MINECO under Grants FIS2017-87970-R and EQC2018-004117-P, and the European Union's Horizon 2020 research and innovation programme under the Marie Sklodowska-Curie Individual Fellowship grant agreement No. 798264.